\documentclass[aps,prl,twocolumn,showpacs,graphicx,nofootinbib]{revtex4}

\usepackage{amsmath,amssymb,graphics,graphicx,bbm}

\usepackage{graphics,graphpap}
\usepackage{graphicx}
\usepackage{epsfig}
\usepackage{epstopdf}
\def\roughly#1{\mathrel{\raise.3ex\hbox
{$#1$\kern-.75em\lower1ex\hbox{$\sim$}}}}

\newcommand{\pslash}{D\kern-0.15em\raise0.17ex\llap{/}\kern0.15em\relax}

\begin{document}

\title{Establishing Analogies between the Physics of Extra Dimensions and Carbon Nanotubes}

\author{Jonas de Woul$^a$\footnote{\tt jodw02@kth.se}, Alexander Merle$^{a,b}$\footnote{{\tt A.Merle@soton.ac.uk} (corresponding author)}, and Tommy Ohlsson$^a$\footnote{\tt tommy@theophys.kth.se}}

\affiliation{$^a$Department of Theoretical Physics, School of Engineering Sciences, KTH Royal Institute of Technology -- AlbaNova University Center, Roslagstullsbacken 21, 106 91 Stockholm, Sweden\\
$^b$ Physics and Astronomy, University of Southampton, Highfield, Southampton SO17 1BJ, United Kingdom
}

\date{\today}
\begin{abstract}
We point out a conceptual analogy between the physics of extra spatial dimensions and the physics of carbon nanotubes which arises for principle reasons, although the corresponding energy scales are at least ten orders of magnitude apart. For low energies, one can apply the Kaluza--Klein description to both types of systems, leading to two completely different but consistent interpretations of the underlying physics. In particular, we discuss in detail the Kaluza--Klein description of armchair and zig-zag carbon nanotubes. Furthermore, we describe how certain experimental results for carbon nanotubes could be re-interpreted in terms of the Kaluza--Klein description. Finally, we present ideas for new measurements that could allow to probe concepts of models with extra spatial dimensions in table-top experiments, providing further links between condensed matter and particle physics.
\end{abstract}

\pacs{04.50.Cd, 14.80.Rt, 61.48.De}
\maketitle

Among the most fascinating ideas in physics are those that question aspects we are so much used to that we nearly take them for granted. One example, questioning the very nature of space-time, is the possible existence of extra spatial dimensions (EDs) in addition to the three we experience in our daily life. Probing the existence of EDs is one of the main tasks of the Large Hadron Collider (LHC), and the corresponding models are generally referred to as \emph{Kaluza--Klein (KK) theories}. Depending on the nature of the additional spatial dimensions, one speaks of \emph{universal extra dimensions}~\cite{ACD}, \emph{large extra dimensions}~\cite{ADD}, or \emph{warped extra dimensions}~\cite{RS}. Indeed, there are recent limits on the energy scale of, e.g., universal extra dimensions (UEDs) from LHC~\cite{LHC}, which amount to a limit on the inverse compactification radius of roughly $\hbar c/R\gtrsim 700$~GeV, i.e., $R\lesssim 0.3\cdot 10^{-18}$~m.

The basic concept of the KK-construction generically used in models with EDs is easy to understand~\cite{ED-Intro}: In case there exists one spatial ED, in addition to the three known ones, this ED must be \emph{compactified} (i.e., ``rolled up'') in order not to modify the law of gravity on large scales~\cite{Hoyle:2004cw}. The easiest possibility is the compactification on a circle $S^1$, leading to periodic boundary conditions of any field along the ED. In this work, we will be interested in applying this to fermionic fields in two spatial dimensions. However, the general construction can, without loss of generality, be exemplified using the more transparent case of a bosonic field in any dimension. To this end, consider the action of a massless real scalar field $\Phi$ in $d+1$ space-time dimensions,
\begin{equation}
 S = \int {\rm d}^d x \int {\rm d}y\ \frac{1}{2} \left( \partial_A \Phi \right) \left( \partial^A \Phi \right),
 \label{eq:5D-action}
\end{equation}
where from now on we mostly use natural units $\hbar=c=1$. Note that the convention is to sum over repeated indices $A=0,1,2,\ldots,d$ in a ($d$+1)-dimensional Minkowski space. A generic example is 3 ordinary spatial dimensions plus 1 temporal dimension and 1 extra spatial dimension, leading to $d+1=5$. Now, the trick is to separate the dynamics in the ED, i.e., $\Phi (x^\mu, y) = \sum_n \phi_n ( x^\mu ) \chi_n (y)$, and to make use of the boundary condition to expand the $y$-dependent part in Fourier modes. While the actual form of these Fourier modes depends very much on the geometry of the ED (e.g.\ sine and cosine functions for compactifying a flat ED on $S^1$), the generic effect on the $d$-dimensional submanifold is the existence of a so-called \emph{KK-tower} of states with increasing masses $m_n$ ($n = 0, 1, 2,\ldots$), as usual for a 1D-potential with certain boundary conditions, their size being characterized by the compactification scale $R$ of the ED. Integrating out the extra spatial dimension in the above example, one obtains
\begin{equation}
 S = \int {\rm d}^d x\ \frac{1}{2} \sum_n \left[ \left( \partial_\mu \phi_n \right) \left( \partial^\mu \phi_n \right) - \frac{n^2}{R^2} \phi^2_n  \right],
 \label{eq:4D-KK}
\end{equation}
leading to the identification of masses $m_n \equiv \frac{n}{R}$. Note that, in particular, the above construction is independent of the actual number of dimensions $d+1$, so instead of using $d=4$, one could ask if there is an easily accessible system in Nature that exhibits an analogous behavior, but for a different number of dimensions.

An illustrative example that has these features is a carbon nanotube (CN)~\cite{CN-Review}, which may be thought of as a two-dimensional sheet of graphite, i.e.\ graphene, rolled up to a thin tube. As is well-known for graphene, after taking the continuum limit, certain low-energy electronic states obey a 2D Dirac-type equation and resemble massless fermions~\cite{Semenoff} (with the Fermi velocity $v_F$ playing the role of the speed of light $c$), similar to our starting point in Eq.~\eqref{eq:5D-action}. This feature allows to study different fundamental concepts of relativistic quantum mechanics, such as the Klein paradox~\cite{Klein-Paradox}, using graphene. If one compactifies 2D continuous space into a cylinder and applies the KK-construction, one naturally obtains a KK-tower of 1D Dirac fermions. This suggests that the low-energy physics of a CN (for which space is instead discrete) allows for a similar interpretation. In this work, we point out that this is indeed true, although the details turn out to be much richer due to the discrete atomic lattice (see e.g.~\cite{Dienes} for a discussion on modular symmetries in carbon nanotori). Note that a CN has a typical length scale set by its radius, which in turn is of the order of the interatomic spacing $a\simeq 2.46$~\r{A}~\cite{Graphene-Review}, and, as we will discuss, this leads to KK-levels at eV-energies. Both CNs and graphene can be manufactured, see~\cite{Tube_Production} for CNs and~\cite{Graphene_Production} for graphene. The production procedure of graphene has been awarded the Nobel Prize in Physics in 2010, and fullerenes, of which a cylindrical form is simply a CN, led to the Nobel Prize in Chemistry in 1996~\cite{Nobel}.

The goal of this work is to emphasize the analogies between KK-theories and the physics of CNs, and to exchange ideas between both fields -- in the same spirit as the Higgs mechanism and the Meissner effect, mass gaps and insulators, etc. This possibility has up to now only been vaguely mentioned (see e.g.~\cite{ModPhys}), but, to our knowledge, never been worked out in any detail. Some of our findings are well-known to condensed matter physicists, but not presented in a way accessible to particle physicists. In turn, a condensed matter physicist could benefit from adopting a new language to describe the phenomena in a CN, which allows for a different way of thinking. Finally, we want to point out possible experiments that could probe the concepts of KK-theories in CN-systems. Actually, some experiments were already performed, but, to our knowledge, have never been interpreted in the context of the KK-description.

To set the stage, we briefly discuss the transition from a graphene sheet to a CN. Extensive descriptions can, e.g., be found in the reviews~\cite{CN-Review,Graphene-Review}, while we here just summarize the basic steps. The typical starting point for graphene is a \emph{tight-binding model} of free electrons (see~\cite{interaction} for cases where interactions are included) on a honeycomb lattice with each triangular sublattice generated by two lattice vectors $\vec{a}_{1,2} = \frac{a}{2} \left( \sqrt{3}, \pm 1 \right)$. The electrons are able to tunnel, or \emph{hop}, between nearest-neighbor sites with an amplitude set by $t\simeq 2.9$~eV, the so-called \emph{hopping strength}. The tight-binding Hamiltonian is diagonalized by transforming to Fourier space. The resulting dispersion relation (one-particle energy eigenvalues) is
\begin{equation}
 E = \pm t \sqrt{3 + 2 \cos ( a k_1 ) + 2 \cos \left[ a ( k_1 - k_2 ) \right] + 2 \cos ( a k_2 )},
 \label{eq:graphene_2}
\end{equation}
with lattice momenta $\vec{k}= \frac{a}{2\pi} ( k_1 \vec{G}_1 + k_2 \vec{G}_2 )$, lattice constant $a\simeq 2.46$~\r{A}, and reciprocal lattice vectors $\vec{G}_{1,2} = \frac{2\pi}{\sqrt{3} a} \left( 1, \pm \sqrt{3} \right)$~\cite{CN-Review,Graphene-Review,Condensed-FT}. At half-filling (one electron per lattice site), the so-called \emph{Fermi surface} in Fourier space consists of just two points $\vec{k}_F^\pm$ [with $E({\vec{k}_F^\pm})=0$]. Expanding Eq.~\eqref{eq:graphene_2} to linear order around these points one finds $E \simeq \pm \frac{3}{2} t |\vec{k}' |$, where $\vec{k}'=\vec{k}-\vec{k}_F^\pm$. This suggests the interpretation of one-particle states close to the Fermi surface as massless Dirac fermions (where the spin quantum number is replaced by a quantity called \emph{quasi-spin}~\cite{Graphene-Review}), with positive and negative energies corresponding to particle and antiparticle states. 

\begin{figure*}[t]
\begin{center}
\begin{tabular}{lr}
\includegraphics[width=8.2cm]{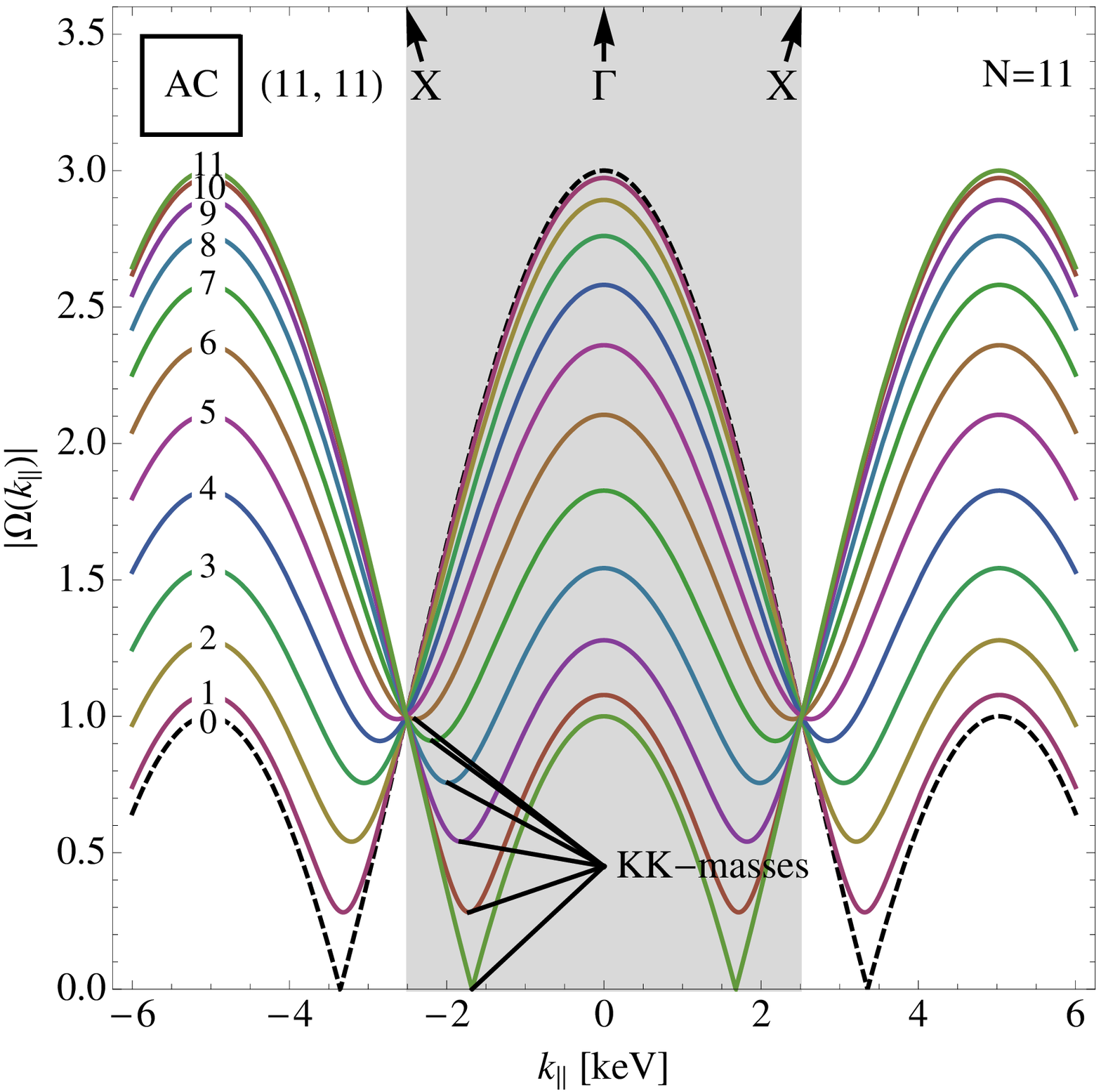} &
\includegraphics[width=8.2cm]{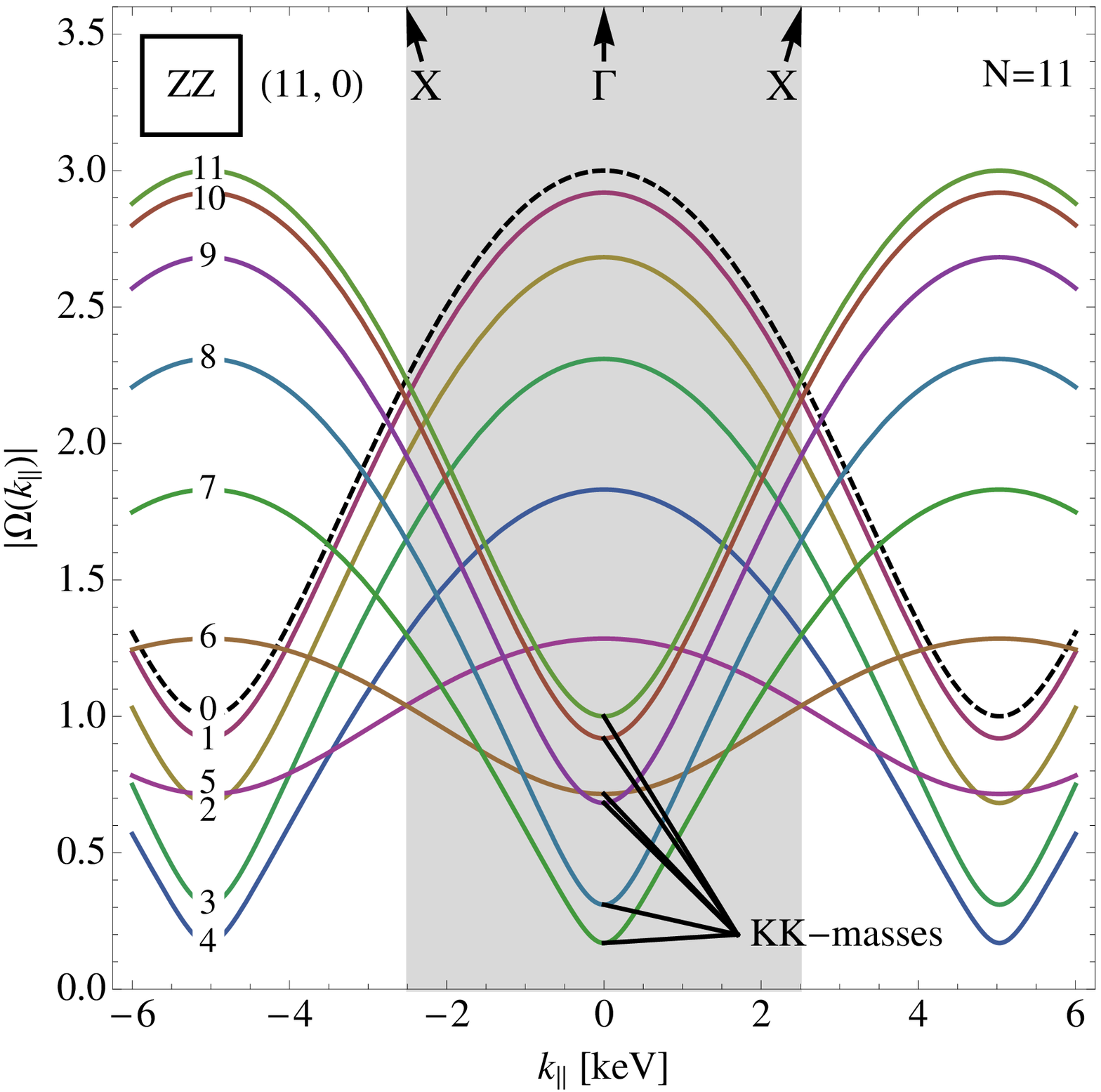}
\end{tabular}
\end{center}
\caption{\label{fig:dispersion} Dispersion relations for the armchair (AC) and zig-zag (ZZ) carbon nanotubes. The negative energy bands, which would correspond to reflections on the $k_\|$-axis, are not displayed. The expansion points and the KK-masses are indicated by the black lines pointing at them. For a precise definition of the $\Gamma$ and X points, see~\cite{CN-Review}.}
\end{figure*}

The transition from the infinite honeycomb lattice to a general (\emph{chiral}) carbon nanotube is achieved by imposing the periodicity condition $\Psi (\vec{r} + N_1 \vec{a}_1 + N_2 \vec{a}_2) = \Psi (\vec{r})$ on any one-particle state, with two integer numbers $(N_1, N_2)$. This leads to the quantization of the \emph{orthogonal} momentum $k_\perp$ along $N_1 \vec{a}_1 + N_2 \vec{a}_2$, while the momentum $k_\|$ parallel to the extension of the CN can be treated as continuous. The illustrative cases, to be discussed here, are the \emph{armchair} $(N,N)$ and \emph{zig-zag} $(N,0)$ CNs. Now, starting from the symmetry of the state, one can derive a quantization condition for $k_\perp$ which translates into a new dispersion relation resulting from Eq.~\eqref{eq:graphene_2}. The results are summarized in Tab.~\ref{tab:compare} and will be discussed in the following. We have also plotted two cases in Fig.~\ref{fig:dispersion}, which exhibit all qualitative features that can appear.

\begin{center}
\begin{table*}[t]
\begin{tabular}{|c||c|c|c|}
\hline
Properties & \multicolumn{2}{|c|}{Carbon nanotubes} & KK-interpretation \\ \cline{2-3}
 & Armchair $(N,N)$ & Zig-zag $(N,0)$ & \\ \hline \hline
Dispersion relation & $\Omega^2(k_\|) = 3+4 c_{1\perp} c_{1\|} + 2 c_{2\|}$ & $\Omega^2(k_\|) = 3+4 c_{1\perp} c_{1\|} + 2 c_{2\perp}$ & Full effect of the ED \\
 $\Omega^2(k_\|) \equiv E^2(k_\|)/t^2$ &  &  &\\ \hline
Quantized momentum & $k_\perp = \frac{2\pi}{a} \frac{n}{N}$ ($-N < n \leq N$) & $k_\perp = \frac{2\pi}{a} \frac{n}{N}$ ($0\leq n \leq N$) & ED-momentum\\ \hline
CN radius $r$ & $\frac{\sqrt{3} m a}{2 \pi}$ & $\frac{m a}{2 \pi}$ & Compactification radius $R$\\ \hline
Minima of the dispersion & $m_n^2 = t^2 \sin^2 \left( \pi \frac{n}{N} \right)$ & $m_n^2 = t^2 \left[ 1 + 2 \cos \left( \pi \frac{n}{N} \right) \right]^2$ & KK-masses\\ 
& $(|n| \geq N/2)$ & $(n> N/2)$ & \\ \hline
Positions of the minima & $k_\|=\frac{2}{a} \arccos \left( - \frac{1}{2} \cos \left( \pi \frac{n}{N} \right) \right)$ & $k_\|=0$ & Zero-momentum states \\ \hline
Number of non-deg.\ minima & $\lfloor  \frac{N+2}{2} \rfloor$ & $\lfloor  \frac{N+1}{2} \rfloor$ & Number of KK-levels \\ \hline
Gapless at half-filling & always & only if $3 | N$ & Existence of a massless mode\\ \hline
\end{tabular}
\caption{\label{tab:compare} Translation between the physics of CNs and KK-theory, and vice versa. We have defined: $c_{l\mathcal{x}} = \cos \left( l \frac{a}{2} k_\mathcal{x} \right),\ l\in \{ 1, 2 \},\ \mathcal{x}\in \{ \|, \perp \}$.}
\end{table*}
\end{center}

After having derived the dispersion relations for the CN under consideration, one can right away observe the behavior resembling a typical KK-theory: The low energy physics is dominated by the minima of the low-lying bands. Around those minima at the positions $k_{\| n}$ one can perform Taylor expansions in $\bar{k}=k_\| - k_{\| n}$, leading to dispersion relations of the form $E^2 = m_n^2 + \bar{k}^2 + \mathcal{O}(\bar{k}^3)$, which can, for small momenta, be interpreted as free particles with masses $m_n$. For example, for a zig-zag CN, all minima are at $k_{\| n}=0$. Expanding the zig-zag dispersion relation from Tab.~\ref{tab:compare} leads to
\begin{equation}
 E^2 = t^2 \left[ 1 + 2 \cos \left( \pi \frac{n}{N} \right) \right]^2 - \frac{t^2}{2} \cos \left( \pi \frac{n}{N} \right) a^2 \bar{k}^2 + \mathcal{O}(\bar{k}^3).
 \label{eq:ZZ-free}
\end{equation}

One can immediately read off that the masses are given by $m_n = t \left[ 1 + 2 \cos \left( \pi \frac{n}{N} \right) \right]$, as long as there is indeed a minimum, which happens for $\cos \left( \pi \frac{n}{N} \right) < 0$ or, equivalently, $n > N/2$. For the example in Fig.~\ref{fig:dispersion}, we have chosen $N=11$, and there should be minima (and hence KK-levels) at $n=6,7,8,9,10,11$ (cf.\ Tab.~\ref{tab:compare}), which is perfectly realized as can be seen from the right panel in the figure. This is particularly interesting in the light of recent LHC bounds by the ATLAS and CMS collaborations~\cite{Nishiwaki:2011gk} limiting the number of KK modes to only ``a few'' in certain models~\cite{Blennow:2011tb}. Note that the KK-masses are \emph{not} equidistant, as is (approximately) the case for typical UED models. The reason for this is the more complicated geometry of the CNs: Space is discrete rather than continuous, so we cannot expect the same behavior as for the simplest particle physics models. In particle physics, it is also known that more complicated geometries lead to more involved KK-spectra, see e.g.~\cite{RS,Burdman:2005sr}.\footnote{Another effect visible is that the number of KK-states is finite, due to the existence of a maximum momentum, translating into a natural cutoff scale of the theory. In this respect, note the similarities to dimensional deconstruction~\cite{deconstruction}.} However, around the regions where the sine and cosine functions in the KK-masses cross zero, one can approximate them by keeping only linear terms, in which case the masses are roughly distributed equidistantly (e.g., $m_n - m_{n+1} \approx \frac{2\pi}{N} t$ for the zig-zag case).\footnote{A physical interpretation of this approximation would be that the momentum is so small that it cannot ``resolve'' the discrete nature of space, and the compactification appears more or less like on a circle. In this limit, the states should also resemble an approximate conservation of KK-number/KK-parity.}

Note that, contrary to typical examples in particle physics, the \emph{zero mode}, i.e., the state with $n=0$, is not the one with the lowest mass. This is a further effect of the non-trivial geometry. However, there always exists a mode with lowest mass. This smallest mass may or may not be zero (cf.\ Tab.~\ref{tab:compare}), depending on the exact values of the parameters. In case there is a massless mode, this means that there is no gap between the positive and negative energy states, which, as is well-known in condensed matter physics, signals that the CN exhibits metallic behavior. If the smallest mass is non-zero, there will be a non-zero gap, making the CN semiconductive rather than metallic (cf.~\cite{CN-Review}). This is another example of two different, but fully equivalent, interpretations of the same physics.

An important point to observe is the order of magnitude of the KK-masses, which is proportional to the hopping strength $t\simeq 2.9$~eV: Studying Tab.~\ref{tab:compare}, one can see that the typical size of a KK-mass in a CN is $\mathcal{O}(1)$~eV. This coincidence is particularly interesting, since this size corresponds to the energy of \emph{visible} light, thereby enabling the investigation with ordinary lasers, which are the main tools used in experimental quantum optics. This suggests that it would indeed be possible to probe some of the concepts originally developed for models with EDs in the analogue system of a CN by a table-top experiment, without the need of constructing a complex particle physics experiment.

We want to discuss possibilities for such tests. This could go in two directions: On the one hand, several experiments have already been performed, but there is no interpretation in the language of KK-theories available. On the other hand, the two pictures could also stimulate each other, by probing concepts arising in models with EDs through testing the analogous behavior of CNs.

A key concept used for condensed matter systems is the so-called \emph{density of states} (DOS), representing the number of one-particle states available for a given energy interval. Any points where the dispersion relation has a vanishing derivative will show up in the DOS as so-called \emph{van Hove singularities}~\cite{vHs}. In particular, the positions of the KK-masses will result in such singularities. Hence, by measuring the DOS, one can probe the existence of the whole KK-tower. This measurement has already been performed for certain types of CNs, e.g., using optical absorption and emission~\cite{CN_optics} or Raman scattering techniques~\cite{Raman}. In the measured spectra, one can identify the low-lying van Hove singularities, whose energies exactly correspond to the KK-construction, illustrating that it is indeed possible to apply the KK-picture.

One more example of an effect in the CN that could potentially have an impact on the picture of KK-theories is the following: In 1D, due to the anti-symmetry of the electromagnetic field strength tensor, there is no magnetic field but only an electric field. Now imagine a CN subject to a homogenous magnetic field $B$ (orthogonal to the CN extension), and a (hypothetical) 1D observer living on the CN and not knowing about the existence of the ``extra'' second and third spatial dimensions. If an electron moves along the CN, it would feel a Lorentz force perpendicular to its direction of motion. However, since the momentum in this perpendicular direction is quantized, the Lorentz force can only have an effect if it is strong enough to lift the electron into an excited KK-state: Setting the Lorentz force $e v_F B$ equal to the central force $p^2/(m_e r)$ and taking the characteristic momentum scale to be $p=\hbar/r$, we obtain for the classical magnetic field strength required to induce a KK-transition $B\sim 0.75\ {\rm T} / (r[{\rm nm}])^3$, where the radius $r$ of the CN is measured in units of nm. Since the 1D observer would have no possibility to measure this magnetic field (any charge in the CN could only move in a direction orthogonal to the Lorentz force), and hence to know of the existence of this field, such a transition to a higher KK-mode would look entirely spontaneous for the 1D observer.

Returning to particle physics, there are models which assume certain forces (e.g.\ gravity~\cite{RS}) to be able to propagate in the full-dimensional \emph{bulk}, while the standard model, or a part of it, lives on the three-dimensional \emph{brane} (see e.g.~\cite{Chang:1999nh} for a case where much more than gravity lives in the bulk). Having the example of the 1D observer in mind, one could ask if it might be possible to derive bounds on the strengths of external forces similar to the one described above and on the size (and maybe even on the geometry) of the bulk from the \emph{non-observation} of spontaneous excitations in Nature. Then, using CNs, one could probe analogous effects in the laboratory in order to test the validity of these considerations. Note also the richer structure: For the CN, the ``bulk'' (i.e., the 2D system) is actually embedded in an even higher-dimensional space, namely our usual 3D world. In this context, it is worth mentioning recent work on the Casimir effect in CNs~\cite{Casimir}.

The CN system suggests an easy picture allowing to think about such possibilities, and to perform at least proof-of-principle experiments with low-dimensional systems. Further possible experiments could be imagined, e.g.\ probing the approximate conservation of KK-number for small momenta (by measuring the transition rates of electrons from higher KK-levels) or the exact contribution of the virtual states in the KK-tower to processes like KK-electron/KK-hole pair annihilation (by comparing the measured de-excitation rates with the calculated annihilation rates using Feynman diagram techniques).

In conclusion, the purpose of this work is to draw the attention of both communities, particle and condensed matter physics, to the illustrative analogies between KK-theories and CN systems. Ideally, one should be able to find more systems (like ordinary graphene or semiconductor junctions, which are close to but certainly not perfectly two-dimensional) where effects of, e.g., non-trivial space-times and/or compactifications could be probed. These considerations could then be extended to more elaborate models exploiting such effects to motivate certain properties that could be useful in particle physics, e.g.\ the existence of a natural cutoff for the KK-tower. Our hope is that this work will inspire subsequent studies of this interdisciplinary topic.

\emph{Acknowledgements:} We would like to thank M.~Abb, M.~D\"urr, and E.~Langmann for useful discussions, and we are particularly grateful to E.~Langmann for valuable comments on the manuscript. This work was supported by the Swedish Research Council (Vetenskapsr\r{a}det) under contract no.\ 621-2011-3985 (TO) and by the G\"oran Gustafsson Foundation (JdW \& AM). AM is now supported by the Marie Curie Intra-European Fellowship within the 7th European Community Framework Programme FP7-PEOPLE-2011-IEF, contract PIEF-GA-2011-297557.

\end{document}